# Homogeneous Focusing Field for Short Relativistic Electron Bunches in Plasma

V.I. Maslov[1,2], I.P. Levchuk[1], D.S. Bondar[1,2], I.N. Onishchenko[1]
[1]NSC Kharkov Institute of Physics & Technology, 61108 Kharkov, Ukraine
[2]Karazin Kharkov National University, Kharkov, Ukraine

Plasma wake lens in which all short relativistic electron bunches of sequence are focused identically and uniformly is studied analytically and by numerical simulation. For two types of lenses necessary parameters of focused sequence of relativistic electron bunches are formulated. Verification of these parameters is performed by numerical simulation.

## 1. INTRODUCTION

As plasma in experiment is inhomogeneous and nonstationary and properties of wakefield changes at increase of its amplitude it is difficult to excite wakefield resonantly by a long sequence of electron bunches (see [1, 2]), to focus sequence (see [3-8]), to prepare sequence from long beam (see [9-11]) and to provide large transformer ratio (see [12-18]). Providing a large transformer ratio is also being studied in dielectric accelerators (see [19-24]). In [2] the mechanism has been found and in [25-29] investigated of resonant plasma wakefield excitation by a nonresonant sequence of short electron bunches.

In [30-33] it has shown that at certain conditions the laser wakefield acceleration is added by a beam-plasma wakefield acceleration.

In [34] point self-injected and accelerated electron bunch was observed.

In numerical simulation [33] on wakefield excitation by a x-ray laser pulse in a metallic-density electron plasma the accelerating gradient of several TV/m has been obtained.

Focusing of relativistic electron bunches of sequence by wakefield, excited in the plasma, is important and interesting (see [35, 36-46]). Mechanism of focusing in the plasma, in which all bunches of sequence are focused identically and uniformly, is proposed and investigated by numerical simulation in [3, 4]. This plasma wake lens for short relativistic electron bunches is studied in this paper analytically and by numerical simulation by code lcode [47]. Unbounded nonmagnetized homogeneous plasma is considered. The rectangular in longitudinal direction bunches (i.e. the bunch current is const along bunch axis) are considered in the fixed their current approximation.

The purpose of this paper is to show analytically and by numerical simulation that one can derive and achieve conditions of identical and uniform focusing of sequence of short relativistic electron bunches.

## 2. HOMOGENEOUS FOCUSING FIELD FOR SEQUENCE OF SHORT BUNCHES

Plasma wake lens with a uniform focusing force for sequence of bunches, whose lengths are equal to half of the excited wavelength $\xi_b=\lambda/2$, with the first bunch, whose charge is in 2 times less than the charges of other bunches $Q_1=Q_i/2$, i=2, 3, ... , the space interval between bunches equals $\lambda$, has been numerically simulated in [3, 4]. We consider a homogeneous focusing of sequence of short bunches.

At first we will derive the wakefield inside the 2nd bunch, the charge of which is more in B times than the charge of the first bunch. We take into account that between the newly generated wakefield and wakefield, formed by previous bunch, is the phase difference A. Then in the second rectangular in longitudinal direction bunch (i.e. the bunch current is const along bunch axis) we have

$$E_z \propto Z_z(\xi) = (1/k)\sqrt{2}\cos(k\xi-\pi/4+A) + B\int_0^\xi d\xi_0 \cos[k(\xi-\xi_0)] =$$
$$= (1/k)\{\sqrt{2}\cos(k\xi-\pi/4+A)+B\sin(k\xi)\}. \quad (1)$$

$Z_z(\xi)$ becomes zero, if $B=\sqrt{2}$ and $A=3\pi/4$.

The same $Z_z(\xi)$ is occurred in each of the following bunches, if they are spaced apart by a $k\delta\xi=2\pi$.

Now we will derive the wakefield inside the 2nd short bunch, $\xi_b=\lambda/4$, charge of which is in 2 times more than the charge of 1st bunch, the space interval between it and the first bunch is equal to $\delta\xi=\lambda$. We take into account that between the newly generated wakefield and wakefield, formed by previous bunch, is the phase difference $\pi$. Then in the 2nd bunch we have

$$E_z \propto Z_z(\xi) = (2/k)\sin(x+\pi) + 2\int_0^\xi d\xi_0 \cos[k(\xi-\xi_0)] = 0, \quad (2)$$
$$E_r \propto Z_r(\xi) = -(2/k)\cos(x+\pi) + 2\int_0^\xi d\xi_0 \sin[k(\xi-\xi_0)] = 2/k. \quad (3)$$

The same $Z_z(\xi)$ and $Z_r(\xi)$ are occurred inside the all next bunches, which are identical to 2nd bunch. From Fig. 1 and Fig. 2 one can see that $E_z=0$ in the bunch location areas.

From Figs. 1-4 one can see that in the areas of bunch location $F_r$ approximately does not depend on the longitudinal coordinate.

We consider now the focusing fields, which are formed in the areas of location of short bunches of sequence (with a certain bunch - precursor) at shaping of their charge linearly along the sequence as well as along each bunch. From Fig. 5 one can see that the bunches are in small and approximately identical decelerating fields and maximal focusing fields.

We consider now focusing field, which is formed in the area of location of a continuous beam, the front of which is a half-step density of length of the half-wavelength (see Fig. 6).

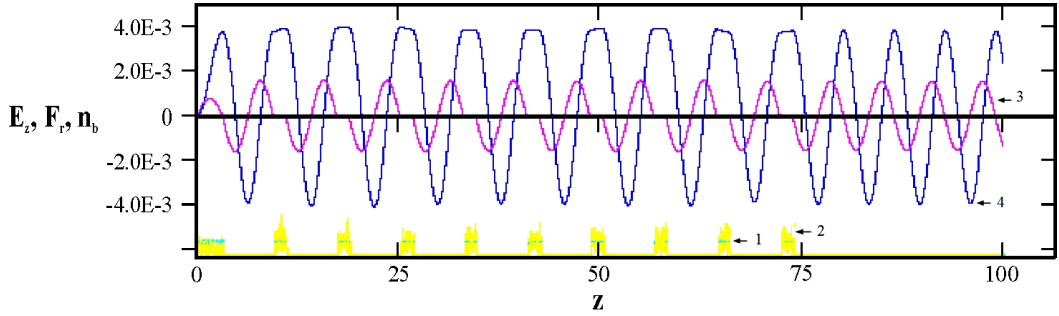

*Fig. 1. The longitudinal distribution of bunch radius (1), density $n_b$ of bunches (2), $E_z$ (3) and $F_r$ (4), excited by sequence of bunches, length of the 1st bunch is equal to $\Delta\xi_{b1}=\lambda/2$, all other bunches are short, $\Delta\xi_b=\lambda/4$, the charge density of all other bunches is in two times larger than the charge of 1st bunch, space intervals between all bunches are equal to $\delta\xi=\lambda$. The longitudinal coordinate z is normalized on $2\pi/\lambda$. $E_z$ and $F_r$ are normalized on $mc\omega_{pe}/e$. e, m are the charge and mass of the electron, c is the light velocity, $\omega_{pe}$ is the electron plasma frequency*

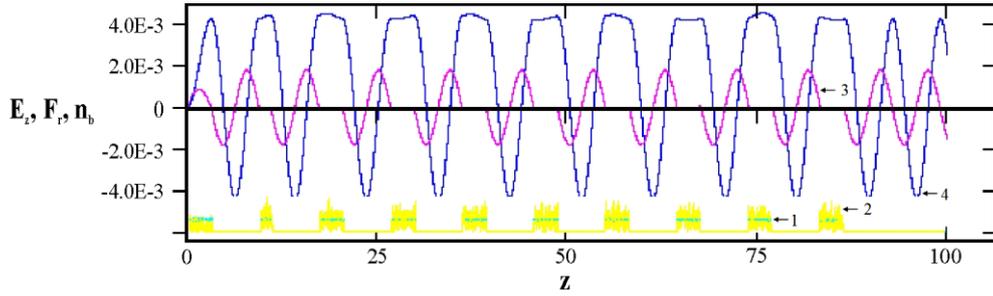

*Fig. 2. The longitudinal distribution of bunch radius (1), density $n_b$ of bunches (2), $E_z$ (3) and $F_r$ (4), excited by sequence of bunches, the length of 2nd bunch is equal to $\Delta\xi_{b1}=\lambda/4$, lengths of the other bunches are $\Delta\xi_{b1}=\lambda/2$, the charge density of all other bunches is in two times larger than the charge density of 1st bunch, the space intervals between all bunches are equal to $\delta\xi=\lambda$*

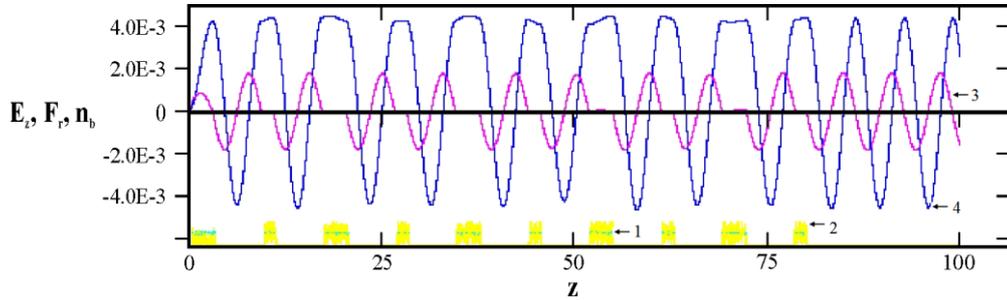

*Fig. 3. The longitudinal distribution of bunch radius (1), density $n_b$ of bunches (2), $E_z$ (3) and $F_r$ (4), excited by sequence of bunches, the length of even bunches is equal to $\Delta\xi_{b2}=\lambda/4$, and the length of uneven bunches is equal to $\Delta\xi_b=\lambda/2$, the charge density of all other bunches is in two times larger than the charge density of 1st bunch, the space intervals between all bunches are equal to $\delta\xi=\lambda$*

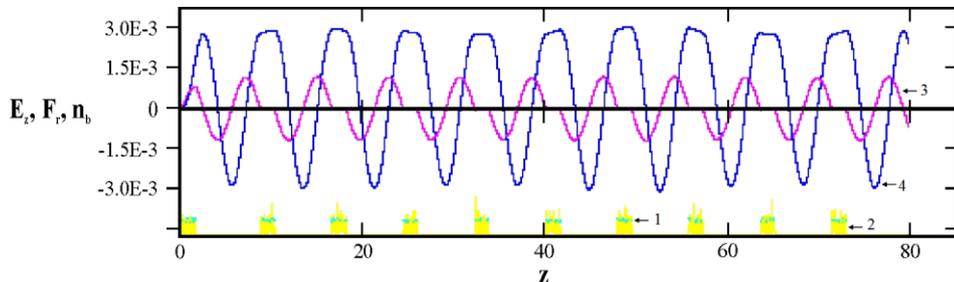

*Fig. 4. The longitudinal distribution of bunch radius (1), density $n_b$ of bunches (2), $E_z$ (3) and $F_r$ (4), excited by sequence of bunches; the charges of bunches are in $\sqrt{2}$ times larger than the charge of 1-st bunch, the space interval between 1-st and 2-nd bunches is equal to $\lambda 9/8$, the space interval between other bunches is equal to $\lambda$*

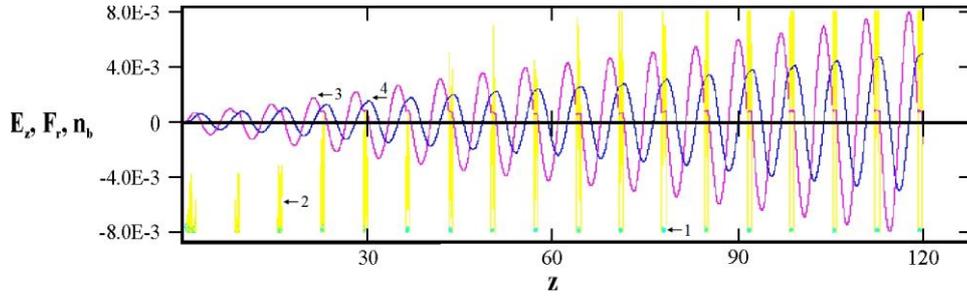

*Fig. 5. The longitudinal distribution of bunch radius (1), density $n_b$ of bunches (2), longitudinal wakefield $E_z$ (3) and focusing field $F_r$ (4), excited by sequence of short bunches (with a certain bunch - precursor) at shaping of their charge linearly along the sequence as well as along each bunch*

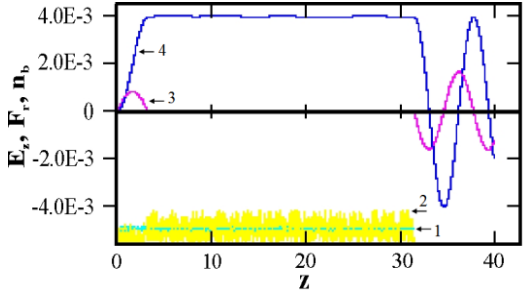

*Fig. 6. The longitudinal distribution of beam radius (1), density $n_b$ of beam (2), longitudinal wakefield $E_z$ (3) and focusing field $F_r$ (4), excited by a continuous beam the front of which is a step*

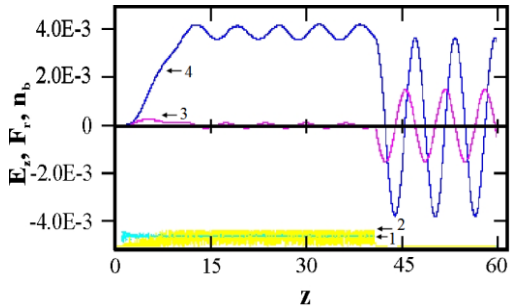

*Fig. 7. The longitudinal distribution of beam radius (1), density $n_b$ of beam (2), longitudinal wakefield $E_z$ (3) and focusing field $F_r$ (4), excited by a continuous beam with a smooth front, the length of which is equal to two wavelengths*

One can see that along the whole beam the focusing field is identical (homogeneous). The focusing field reaches its maximal value at the short spatial interval coinciding with the length of the beam front.

Such a uniform focusing field along the beam is not reached (see Fig. 7) in the case of a smooth beam density increase along beam front.

## 5. CONCLUSION

It has been shown that all bunches of sequences can be focused identically and uniformly. It has been shown that there are two types of lenses for a sequence of short bunches. It is necessary that in one case the length of the 1st bunch $\Delta\xi_{b1}$ should be equal to half of the wavelength $\Delta\xi_{b1}=\lambda/2$, all other bunches are short, $\Delta\xi_b<\lambda/2$, the charge density of all other bunches is in 2 times larger than the charge density of the 1st bunch. The space intervals between all bunches should be multiples of the wavelength $\delta\xi=p\lambda$, p=1, 2, … for achievement of:

-$E_z$=0,

-radial focusing field is the same along the bunch $F_r$=const in areas of location of bunches.

In the second case it is necessary that the charges of all bunches are in $\sqrt{2}$ times larger than the charge of 1st bunch. The space interval between the 1-st and 2-nd bunches is equal to $(n+1/8)\lambda$, n=1, 2, … The space interval between the other bunches is multiple to wavelength. It has been shown analytically that only 1st bunch is in the finite $E_z\neq 0$. Other bunches are in zero longitudinal electric wakefield $E_z$=0. Consequently, the 1-st bunch exchanges by energy with wakefield. Next bunches do not exchange by energy with the wakefield. Wakefield radial force $F_r$ is the same approximately along the bunch. The focusing field of this value is formed in now widely investigated plasma lens for a long relativistic electron beam.